\documentclass[aps,prd,preprint,groupedaddress,showpacs]{revtex4-1}
\def\lvec#1{\setbox0=\hbox{$#1$}
    \setbox1=\hbox{$\scriptstyle\leftarrow$}
    #1\kern-\wd0\smash{
    \raise\ht0\hbox{$\raise1pt\hbox{$\scriptstyle\leftarrow$}$}}
    \kern-\wd1\kern\wd0}
\def\rvec#1{\setbox0=\hbox{$#1$}
    \setbox1=\hbox{$\scriptstyle\rightarrow$}
    #1\kern-\wd0\smash{
    \raise\ht0\hbox{$\raise1pt\hbox{$\scriptstyle\rightarrow$}$}}
    \kern-\wd1\kern\wd0}
\def\diracstar#1#2{
    \setbox0=\hbox{$\gamma$}\setbox1=\hbox{$\gamma_{#1}$}
    \gamma_{#1}\kern-\wd1\kern\wd0
    \smash{\raise4.5pt\hbox{$\scriptstyle#2$}}}
\def\tr{\,\hbox{tr}\,}

\newcommand{\beq}{\begin{equation}}
\newcommand{\eeq}{\end{equation}}
\newcommand{\beqn}{\begin{eqnarray}}
\newcommand{\eeqn}{\end{eqnarray}}

\begin{document}

%\preprint{}

\title{\bf $q \bar q$-potential}


\author{G.C.~Rossi}
\email[]{rossig@roma2.infn.it}
%\thanks{}
\affiliation{Dipartimento di Fisica, Universit\`a di Roma {\it Tor Vergata}\\ \small and INFN, Sezione di Roma 2 \\ \small Via della Ricerca Scientifica, 00133 Roma, Italy\\ \small}

\author{M.~Testa}
\email[]{massimo.testa@roma1.infn.it}
%\thanks{}
\affiliation{Dipartimento di Fisica, Universit\`a di Roma {\it La Sapienza}\\ \small and 
INFN, Sezione di Roma 1 \\ \small P.le A.\ Moro 5, 00185 Roma, Italy}


%\date{}

\begin{abstract}
We show how to define and compute in a non-perturbative way the potential between $q$ and $\bar q$ colour sources in the singlet and octet (adjoint) representation of the colour group.
\end{abstract}

%\keywords{}

\maketitle

\section{Introduction}
\label{sec:INTRO}

There is some interest in computing the potential between two point-like sources with the $q\bar q$ colour quantum numbers in the octet (adjoint) representation of the colour group~\cite{Brown:1979ya,McLerran:1981pb,Nadkarni:1986as,Bodwin:1994jh,Kaczmarek:2002mc,Philipsen:2002az,Jahn:2004qr,Shuryak:2004tx}. However, conflicting  results have been reported in the literature~\cite{Philipsen:2002az,Jahn:2004qr} and the question seems to still need some clarification. 

In this paper we reanalyze the problem discussing, in the pure Yang-Mills (YM) theory, the structure of energy eigenstates in the presence of colour sources and derive explicit formulae expressing the singlet and adjoint potential in terms of the Feynman propagation kernel (sometimes also called the Schr\"odinger functional) computed in the temporal ($A_0=0$) gauge. The theoretical framework we shall use is the formulation of the temporal gauge developed in refs.~\cite{Rossi:1979jf,Rossi:1980pg,Rossi:1982ag}. 

In sect.~\ref{sec:GFOR} we review the general formalism. In sect.~\ref{sec:GLOB} we illustrate the constraints imposed by global colour rotations on the structure of the functional integral. The solution to the problem of characterizing energy eigenstates belonging to different colour representations is given in sect.~\ref{sec:EXC} in the formal continuum theory. In sect.~\ref{sec:SINOCT} we give explicit formulae for extracting the singlet and adjoint $q \bar q$-potential from the knowledge of the Feynman kernel. In sect.~\ref{sec:LTS} we illustrate how to reformulate the previous analysis in the lattice language, suitable for numerical simulations. A number of technical issues are discussed in the Appendices. In Appendix~\ref{sec:APPA} we recall some relevant group theoretical formulae. In Appendix~\ref{sec:APPB} we construct the projectors necessary to single out states with definite global colour transformation properties. In Appendix~\ref{sec:APPC} we recall the formulae necessary to decompose the $[N_c]\otimes [\bar N_c]$ tensor product into the sum of singlet and adjoint representation. For completeness we give in Appendix~\ref{sec:APPD} the expression that the Feynman kernel in the presence of $q\bar q$-sources takes in the Coulomb gauge.

\section{General formalism}
\label{sec:GFOR}

The Feynman kernel in the presence of colour point-like sources belonging to arbitrary colour representations takes in the temporal gauge the expression~\cite{Rossi:1979jf,Rossi:1980pg,Rossi:1982ag}
\beqn
\hspace{-.7cm}&&K({\bf A}_2,\{u_2\};{\bf A}_1,\{u_1\};T)=\int_{{\cal G}_0}{\cal D}\mu(h) {\cal R}(U_h)_{\{u_2\}\{u_1\}}
\widetilde K({\bf A}_2^{U_h},{\bf A}_1;T)\, ,\label{KQQB}
\eeqn
where ${\cal D}\mu(h)$ is the Haar invariant measure over the gauge group, ${\cal G}_0$, of the topologically trivial, time-independent gauge transformation that tend to the identity at spatial infinity. In eq.~(\ref{KQQB}) we have defined   
\beqn
\hspace{-.7cm}&&\widetilde K({\bf A}_2,{\bf A}_1;T)=\int^{{\bf A}({\bf x},T_2)={\bf A}_2({\bf x})}_{{\bf A}({\bf x},T_1)={\bf A}_1({\bf x})} {\cal D}{\bf A}\exp{[-S_{YM}({\bf A}, A_0=0)]}\label{KT}
\eeqn
where $S_{YM}({\bf A}, A_0=0)$ is the YM action taken at $A_0=0$ and 
\beq
{\cal D}{\bf A}=\prod_{{\bf x}, T_1<t<T_2}d{\bf A}({\bf x},t)\, .\label{AMEAS}
\eeq
In eq.~(\ref{KQQB}) we have set $T=T_2-T_1$ and made use of the definitions~\cite{note1}
\beqn
&&A_k^{U_h}({\bf x})=U^\dagger_h A_k U_h ({\bf x})+iU_h^\dagger \partial_k U_h({\bf x}) \, ,\qquad  k=1,2,3\, , \label{gaugetrans}\\
&&U_h({\bf x})= \exp{[i\lambda^ah^a({\bf x})]}\in {\cal G}_0\, ,
\eeqn
where the matrices $\lambda^a$ ($a=1,2,\ldots,N^2_c-1$) are the hermitian SU($N_c$) generators in the fundamental representation $[N_c]$, normalized to $\tr [\lambda^a\lambda^b]=\delta^{ab}/2$ and we have indicated with $\bar \lambda^a=-(\lambda^a)^\star$ the generators of the conjugate fundamental representation $[\bar N_c]$. Finally, we have introduced the compact notation
\beq
{\cal R}(U_h)_{\{u_2\}\{u_1\}}=\prod_{j=1}^{L}\Big{[}\exp{i\lambda^{[j]}_ah_a({\bf x}_j)}\Big{]}_{u_2^j u_1^j}\label{TENS}
\eeq
to represent a set of $L$ sources in the colour representations $[j]$ localized at the points ${\bf x}_j$. In eq.~(\ref{TENS}) we denoted with $[\lambda^{[j]}_a]_{u_2^j u_1^j}$ the SU($N_c$) generators in the representation $[j]$. The indices $u_1^j$ and $u_2^j$ run over values appropriate for the representation $[j]$. 

Since $K({\bf A}_2,\{u_2\};{\bf A}_1,\{u_1\};T)$ is the matrix element of the (Euclidean) time translation operator $\exp(-{\cal H}T)$, we can write its spectral decomposition in the form
\beqn
\hspace{-.7cm}&&K({\bf A}_2,\{u_2\};{\bf A}_1,\{u_1\};T)=\sum_{k}e^{-E_kT}\psi_{k}({\bf A}_2,\{u_2\})\,\psi^\star_{k}({\bf A}_1,\{u_1\})\, ,\label{KEXP}
\eeqn
where the functional $\psi_{k}({\bf A};\{u\})$ is the eigenstate of the Hamiltonian, $\cal H$, corresponding to the eigenvalue $E_k$
\beq
{\cal H}\psi_{k}({\bf A},\{u\})=E_k\psi_{k}({\bf A},\{u\})\label{EIGEN}\, .
\eeq
The state functional $\psi_{k}({\bf A};\{u\})$ transforms, under time independent gauge transformations ($U_w\in {\cal G}_0$), as 
\beq
\psi_{k}({\bf A}^{U_w},\{u\})=\sum_{\{u'\}}{\cal R}^\dagger(U_w)_{\{u\}\{u'\}}\psi_{k}({\bf A},\{u'\})\, .\label{GAUSS}
\eeq
Eq.~(\ref{GAUSS}) is equivalent to the Gauss' law in the Hilbert space sector of states with $L$ external colour sources, as specified above. This can be checked by taking the functional derivative of both sides of the equation with respect to $w_a({\bf x})$ at $w_a({\bf x})=0$.

It was shown in~\cite{Rossi:1979jf} that the kernel $K({\bf A}_2,\{u_2\};{\bf A}_1,\{u_1\};T)$ and the eigenfunctionals $\psi_{k}({\bf A};\{u\})$ are gauge independent in the sense that, had we chosen to quantize the theory in a different canonical gauge, say $F({\bf A})=0$ (typically $F({\bf A})={\bf \nabla} \cdot {\bf A}$), the corresponding kernel and the related eigenfunctionals in the $F$-gauge would coincide with those in the temporal one, when evaluated on a gauge field ${\bf A}$ satisfying $F({\bf A})=0$.

Since the quantity $\sum_{\{u\}} \psi^*({\bf A},\{u\}) \phi ({\bf A},\{u\})$ is invariant under gauge transformations (see eq.~(\ref{gaugetrans})), in order to define a scalar product we must make use of the Faddev--Popov (FP) procedure. This is done by choosing any spatial gauge-fixing condition, say $F({\bf A})=0$, and accordingly defining the scalar product by means of the formula~\cite{Rossi:1983hr}
\beq
(\psi,\phi) \equiv \int {\cal D}\mu_F({\bf A})\sum_{\{u\}} \psi^*({\bf A},\{u\}) \phi ({\bf A},\{u\}) \, ,
\label{NORM}
\eeq
where 
\beq
{\cal D}\mu_F({\bf A})=\prod_{{\bf x}}\Delta_{F}({\bf A}) \delta[F({\bf A})] \,d{\bf A}({\bf x})
\label{MEAS}
\eeq
is the gauge field integration measure. In eq.~(\ref{MEAS}) $\Delta_{F}({\bf A})$ is the FP determinant~\cite{FP} defined by 
\beq
\Delta_{F}({\bf A})\int_{{\cal G}_0} {\cal D}\mu(h) \delta[F({\bf A}^{U_h})]=1\, .
\label{FP}
\eeq
We recall that the value of scalar product in eq.~(\ref{NORM}) is independent on the gauge function $F({\bf A})$ chosen to define it. 

Denoting with $d_k$ the degeneracy of the energy level $E_k$, we can compute the complete trace of the Feynman propagation kernel~(\ref{KEXP}) obtaining 
\beqn
\hspace{-.7cm}&&\int {\cal D}\mu_F({\bf A})\sum_{\{u\}}  K({\bf A},\{u\};{\bf A},\{u\};T)=\sum_{k} d_k e^{-E_kT} \, . \label{tottrace}
\eeqn
The gauge invariance of the l.h.s.\ of eq.~(\ref{tottrace}) implies the gauge invariance of the quantities $d_k$ and $E_k$.

\section{Global colour transformations} 
\label{sec:GLOB}

In this section we discuss how global colour transformations are implemented in the temporal gauge when external sources are present.

Let us denote by $V$ a global colour rotation, i.e.\ a constant SU$(N_c)$ transformation. The global colour invariance of $S_{YM}({\bf A}, A_0=0)$ implies
\beqn
\widetilde K({\bf A}_2^{V},{\bf A}_1^{V};T) = \widetilde K({\bf A}_2,{\bf A}_1;T) \, .
\eeqn
From the definition~(\ref{KQQB}) we then have 
\beqn
\hspace{-.7cm}&&K({\bf A}_2^V,\{u_2\};{\bf A}_1^V,\{u_1\};T)=\nonumber\\
\hspace{-.7cm}&&=\int_{{\cal G}_0}{\cal D}\mu(h) {\cal R}(U_h)_{\{u_2\}\{u_1\}}
\widetilde K({\bf A}_2^{VU_h},{\bf A}_1^V;T)=\nonumber\\
\hspace{-.7cm}&&=\int_{{\cal G}_0}{\cal D}\mu(h) {\cal R}(U_h)_{\{u_2\}\{u_1\}}\widetilde K({\bf A}_2^{VU_hV^\dagger},{\bf A}_1;T)=\nonumber\\
\hspace{-.7cm}&&=\int_{{\cal G}_0}{\cal D}\mu(h) {\cal R}(V^\dagger U_h V)_{\{u_2\}\{u_1\}}\widetilde K({\bf A}_2^{U_h},{\bf A}_1;T)=\nonumber\\
\hspace{-.7cm}&&= {\cal R}(V^\dagger)_{\{u_2\}\{u_2'\}}\int_{{\cal G}_0}{\cal D}\mu(h) {\cal R}(U_h)_{\{u_2'\}\{u_1'\}}\widetilde K({\bf A}_2^{U_h},{\bf A}_1;T) {\cal R}(V)_{\{u_1'\}\{u_1\}}=\nonumber\\
\hspace{-.7cm}&&={\cal R}(V^\dagger)_{\{u_2\}\{u_2'\}}K({\bf A}_2,\{u_2'\};{\bf A}_1,\{u_1'\};T){\cal R}(V)_{\{u_1'\}\{u_1\}}\, .\label{KQQBV}
\eeqn
Notice that the functional change of variables $VU_hV^\dagger\to U_h$ in the third equality of eq.~(\ref{KQQBV}) is allowed by the fact that $VU_hV^\dagger$ is a (topologically trivial) gauge transformation that tends to unit at spatial infinity and hence belongs to ${\cal G}_0$. 

Eq.~(\ref{KQQBV}) can be rewritten in the form 
\beqn
\hspace{-.7cm}&&{\cal R}(V)_{\{u_2\}\{u_2'\}}K({\bf A}_2^V,\{u_2'\};{\bf A}_1,\{u_1\};T)=\nonumber \\
\hspace{-.7cm}&&=K({\bf A}_2,\{u_2\};{\bf A}_1^{V^\dagger},\{u_1'\};T){\cal R}(V)_{\{u_1'\}\{u_1\}}\, .\label{KQQBVR}
\eeqn
Eq.~(\ref{KQQBVR}) allows us to define the operator ${\cal U}(V)$, implementing global colour transformations, which acts on wave functionals as 
\beq
[{\cal U}(V) \psi]({\bf A},\{u\}) ={\cal R}(V)_{\{u\}\{u'\}}\psi({\bf A}^V,\{u'\}) \, .
\label{UNIT}
\eeq
${\cal U}(V)$ provides a unitary representation of the SU($N_c$) group in the Hilbert space of state functionals and commutes with the kernel $K$. 

The unitarity of ${\cal U}(V)$ follows from the chain 
\beqn
\hspace{-.8cm}&&({\cal U}^\dagger ({V})\psi, \phi) = (\psi, {\cal U}(V) \phi)= \int {\cal D}\mu_F({\bf A}) \psi^*({\bf A},\{u'\}) {\cal R} (V)_{\{u'\}\{u\}} \phi ({\bf A}^{V},\{u\}) = \nonumber\\
\hspace{-.8cm}&&=\int {\cal D}\mu_F({\bf A})  [ {\cal R}^\dagger ({V})_{\{u\}\{u'\}}  \psi({\bf A}^{{V}^\dagger},\{u'\}) ]^* \phi ({\bf A},\{u\})\, .
\label{SCALP}
\eeqn
In fact, eq.~(\ref{SCALP}) shows that
\beq
[{\cal U}^\dagger(V)\psi]({\bf A},\{u\})={\cal R}^\dagger(V)_{\{u\}\{u'\}} \psi({\bf A}^{V^\dagger},\{u'\})\, ,
\label{PRO}
\eeq
which, together with eq.~(\ref{UNIT}), gives ${\cal U}(V^\dagger)={\cal U}^\dagger(V)$ entailing the unitarity of ${\cal U}(V)$. 

To prove that the Feynman kernel commutes with ${\cal U}$ we take an arbitrary wave functional, $\psi({\bf A},\{u\})$, in the appropriate source sector (see eq.~(\ref{GAUSS})), and consider the chain of equalities 
\beqn
&&{\cal U}(V)\int {\cal D}\mu_F({\bf A}_1) K({\bf A}_2,\{u_2\};{\bf A}_1,\{u_1\};T)\psi({\bf A}_1,\{u_1\})=\nonumber\\
&&={\cal R}(V)_{\{u_2\}\{u_2'\}}\int {\cal D}\mu_F({\bf A}_1) K({\bf A}_2^V,\{u_2'\};{\bf A}_1,\{u_1\};T)\psi({\bf A}_1,\{u_1\})=\nonumber\\
&&=\int {\cal D}\mu_F({\bf A}_1) K({\bf A}_2,\{u_2\};{\bf A}_1^{V^\dagger},\{u_1'\};T){\cal R}(V)_{\{u_1'\}\{u_1\}}\psi({\bf A}_1,\{u_1\})=\nonumber\\
&&=\int {\cal D}\mu_F({\bf A}_1) K({\bf A}_2,\{u_2\};{\bf A}_1,\{u_1'\};T){\cal R}(V)_{\{u_1'\}\{u_1\}}\psi({\bf A}_1^V,\{u_1\})=\nonumber\\
&&=\int {\cal D}\mu_F({\bf A}_1) K({\bf A}_2,\{u_2\};{\bf A}_1,\{u_1\};T)\,[{\cal U}(V)\psi]({\bf A}_1,\{u_1\})\, ,
\label{COM}
\eeqn
where the third equality is a consequence of the invariance of the measure~(\ref{MEAS}) under global colour rotations. Owing to the arbitrariness of $\psi({\bf A},\{u\})$, eq.~(\ref{COM}) implies that ${\cal U}(V)$ commutes with the Feynman propagation kernel.

Setting ${\bf A}_1={\bf A}_2={\bf A}$ in eq.~(\ref{KQQBV}) and integrating over ${\bf A}$ with the measure~(\ref{MEAS}), we get the key formula
\beqn
&&K_{\{u\}\{u'\}}(T) \equiv \int {\cal D}\mu_F({\bf A}) K({\bf A},\{u\};{\bf A},\{u'\};T)=\int {\cal D}\mu_F({\bf A}) K({\bf A}^V,\{u\};{\bf A}^V,\{u'\};T) =  \nonumber \\
&&={\cal R}(V^\dagger)_{\{u\}\{v\}} K_{\{v\}\{v'\}}(T) {\cal R}(V)_{\{v'\}\{u'\}}\, .\label{KQQBV1}
\eeqn
Eq.~(\ref{KQQBV1}) tells us that $K(T)$ commutes with ${\cal R}(V)$, so that by Schur's lemma it is a multiple of the identity matrix within any irreducible colour source representation. If we are in a sector in which more than one colour source is present, ${\cal R}(V)$ can be decomposed into the direct sum of irreducible representations of SU($N_c$), and $K(T)$ itself is a direct sum of multiples of the unit matrix, one for each irreducible component. We will make use of this result below. 

We end this section with some considerations on what happens if the gauge integration over ${\cal G}_0$ is extended to the set $\overline {\cal G}_0$ of transformations that also includes global colour rotations. In this case the colour averaged kernel 
\beqn
\hspace{-.7cm}&&\overline K({\bf A}_2,\{u_2\};{\bf A}_1,\{u_1\};T)=
\int_{\overline {\cal G}_0}{\cal D}\mu(h) {\cal R}(U_h)_{\{u_2\}\{u_1\}}
\widetilde K({\bf A}_2^{U_h},{\bf A}_1;T)\label{KQQBADB}
\eeqn 
satisfies the property 
\beq
\overline K({\bf A}_2^V,\{u_2\};{\bf A}_1,\{u_1\};T)=
{\cal R}(V^\dagger)_{\{u_2\}\{u_2'\}}\overline K({\bf A}_2,\{u_2'\};{\bf A}_1,\{u_1\};T)\, ,
\label{KQQBADBI}
\eeq 
that implies for the averaged state functionals, $\overline\psi({\bf A},\{u\})$, appearing in its spectral decomposition, the invariance property 
\beq
[{\cal U}(V)\overline\psi]({\bf A},\{u\})={\cal R}(V)_{\{u\}\{u'\}}\overline \psi({\bf A}^V,\{u'\})=\overline \psi({\bf A},\{u\})\, .
\label{INVACS}
\eeq
Eq.~(\ref{INVACS}) means that, for the averaged state functionals, a global colour rotation on the gauge field amounts to a rotation ${\cal R}(V)$ acting only on source indices, or in other words that any such state functional is a global colour singlet. 

\section{The $q\bar q$-source system}
\label{sec:EXC}

In this section we specialize the formulation of sects.~\ref{sec:GFOR} and~\ref{sec:GLOB} to the particular case of two sources with the $q$ and $\bar q$ colour quantum numbers and show explicitly how to define and extract the $q\bar q$-potential in the singlet and adjoint representation. 
 
The Feynman kernel in this particular source sector has the form~\cite{Rossi:1979jf,Rossi:1980pg,Rossi:1982ag}
\beqn
\hspace{-1.2cm}&&K({\bf A_2},s_2,r_2;{\bf A_1},s_1,r_1;T)=\nonumber\\
\hspace{-1.2cm}&&=\int_{{\cal G}_0}{\cal D}\mu(h) \Big{[}\exp{[i\lambda^a h^a({\bf x}_q)]}\Big{]}_{s_2s_1}\Big{[}\exp{[i\bar\lambda^a h^a({\bf x}_{\bar q})]}\Big{]}_{r_2r_1}
\widetilde K({\bf A_2}^{U_h},{\bf A_1};T)=\nonumber\\
\hspace{-1.2cm}&&=\int_{{\cal G}_0}{\cal D}\mu(h) \Big{[}\exp{[i\lambda^a h^a({\bf x}_q)]}\Big{]}_{s_2s_1}\Big{[}\exp{[-i\lambda^a h^a({\bf x}_{\bar q})]}\Big{]}_{r_1r_2}
\widetilde K({\bf A_2}^{U_h},{\bf A_1};T)\, .\label{KQQBAD}
\eeqn
In order to pick up the energies of the lowest lying states one has to study the large $T$ behaviour of the expression~(\ref{KQQBAD}) while, at the same time, projecting out the desired colour structure. To this end one needs to classify the energy eigenfunctionals in terms of their global colour transformation properties.

\subsection{Classification of the energy eigenstates} 
\label{sec:classst}

Since, according to eq.~(\ref{COM}), the Feynman kernel commutes with ${\cal U}(V)$, the energy eigenstates are classified in terms of the irreducible representations of the colour group SU($N_c$).

Every energy eigenstate in $q {\bar q}$-sector is described by a wave functional which can be parametrized as 
\begin{eqnarray}
\psi({\bf A}; s,r) = [\phi ({\bf A}) I + \lambda^a \phi_a ({\bf A})]_{sr} \equiv \phi ({\bf A}) I + \lambda^a \phi_a ({\bf A}) \equiv \psi({\bf A}) \, , \label{wfunc}
\end{eqnarray}
in terms of the $1 + (N_c^2-1)$ functionals $\phi ({\bf A})$ and $\phi_a ({\bf A})$. The parametrization given in eq.~(\ref{wfunc}) holds for any one of the components of an irreducible multiplet. The totality of the members of the multiplet can be reached through global colour transformations, according to the formula
\begin{eqnarray}
{\cal U}(V) \psi({\bf A}) \equiv \psi^V({\bf A}) =V \psi({\bf A}^V) V^\dagger = \phi ({\bf A}^V) I + V \lambda^a V^\dagger \phi_a ({\bf A}^V) \, . \label{globtr}
\end{eqnarray}
As it can be seen from eq.~(\ref{globtr}), the global colour transformation of a state is composed of two different contributions:
\begin{itemize}
\item the ''colour-spin'' contribution coming from the action of the matrix $V$ on the source indices 
\item and the ''orbital'' contribution coming from the transformation ${\bf A} \rightarrow {\bf A}^V$.
\end{itemize}

As we said, any energy eigenstate must belong to an irreducible representation of the colour group, then $\psi({\bf A})$ must span a unique irreducible representation for {\em any} value of $ {\bf A}$, when transformed as in eq.~(\ref{globtr}). Notice, however, that for ${\bf A} = {\bf 0}$ the orbital contribution is suppressed and eq.~(\ref{globtr}) becomes  
\begin{eqnarray}
\psi^V({\bf 0}) =V \psi({\bf 0}) V^\dagger = \phi ({\bf 0}) I + V \lambda^a V^\dagger \phi_a ({\bf 0}) \, . \label{globtr0}
\end{eqnarray}
Eq.~(\ref{globtr0}) implies that the two terms in the r.h.s.\ cannot be simultaneously different from zero, otherwise $\psi^V({\bf 0})$ would belong to the reducible $I \oplus [N_c^2 - 1]$ representation. This means that we have the following three alternatives 

1) $\phi_a ({\bf 0}) =0$ and  $\phi ({\bf 0}) \neq 0$

2) $\phi ({\bf 0}) = 0$ and $\phi_a ({\bf 0}) \neq 0$ (for some $a$) 

3) $\phi ({\bf 0}) = \phi_a ({\bf 0}) =0$ \\
in correspondence to three different types of irreducible representations that we are now going to discuss. 

1) - 2) When $\phi ({\bf 0}) \neq 0$ (or $\phi_a ({\bf 0}) \neq 0$), $\psi({\bf A})$ must belong to the ''colour-spin'' singlet (or adjoint) representation, because the representation it belongs to cannot change discontinuously under a smooth variation of the ``parameters" ${\bf A}$. 

3) For states such that $\phi({\bf 0}) = \phi_a ({\bf 0}) =0$ we can have more general assignments of global colour quantum numbers. 

Recalling that the adjoint representation $R^{[Ad]}(V)$ is defined by
\begin{eqnarray}
V \lambda^a V^\dagger = R_{ca}^{[Ad]} (V) \, \lambda^c \, ,
\end{eqnarray}
with $R^{[Ad]}(V)$ real, eq.~(\ref{globtr}) can be rewritten in the form 
\begin{eqnarray}
\psi^V({\bf A}) = \phi ({\bf A}^V) I + R_{ca}^{[Ad]} (V) \phi_a ({\bf A}^V)  \, \lambda^c \, . \label{transprop}
\end{eqnarray}
The two ${\bf A}$-functionals in the r.h.s.\ of eq.~(\ref{transprop}) must transform under the colour group and therefore will display ``magnetic" quantum numbers. Thus we will denote them as $\phi_m ({\bf A})$, $\phi_{a k} ({\bf A})$ respectively, with transformation properties under ${\bf A}\to{\bf A}^V$ given by 
\begin{eqnarray}
&&\phi_m ({\bf A}^V) = R^{[\alpha]}_{mm'} (V) \phi_{m'} ({\bf A})\, , \label{alpha} \\
&&{\phi_a}_k ({\bf A}^V) = R^{[\beta]}_{kk'} (V) \phi_{a k'} ({\bf A}) \, , \label{beta}
\end{eqnarray}
where $R^{[\alpha]}$ and $R^{[\beta]}$ are SU$(N_c)$ representations. Using eqs.~(\ref{alpha}) and~(\ref{beta}), eq.~(\ref{transprop}) becomes
\begin{eqnarray}
\psi^V({\bf A}) = R^{[\alpha]}_{mm'} (V) {\phi}_{m'} ({\bf A}) I + R_{ca}^{[Ad]} (V) R^{[\beta]}_{kk'} (V) \phi_{a k'} ({\bf A}) \, \lambda^c \, , \label{transprop1}
\end{eqnarray}
showing explicitly that the $\phi_m$'s transform according to representation $[\alpha]$, while the $\phi_{a k}$'s belong to the direct product $[\beta] \otimes [N_c^2-1]$.

Due to the trace-orthogonality of the identity and the $\lambda$-matrices, the two terms in eq.~(\ref{transprop1}) must separately belong to irreducible representations.

We conjecture that the two terms in eq.~(\ref{transprop1}) correspond to two different types of energy eigenfunctionals, although we cannot exclude that, when the two irreducible representations are equivalent, both terms may simultaneously contribute to a given eigenfunctional.

It follows from this analysis that the $[\alpha]$ representation must be irreducible and that, owing to the colour invariance of the dynamics, the ``wave function" $\phi_{a k}$ will be such to single out an irreducible representation from the tensor product $[\beta] \otimes [N_c^2-1]$.

In summary, depending on the structure of the source indices and the colour representation  to which the gluon wave function belongs, we can classify the energy eigenstates in four classes. 
\begin{itemize}

\item Colour-spin singlet, orbital singlet states
\begin{eqnarray}
&&\psi^{[S]}_{[S]} ({\bf A})= \phi ({\bf A}) I \label{colors}\\
&&\phi ({\bf A}^V) = \phi ({\bf A}) \, , \nonumber
\end{eqnarray}
with $\phi ({\bf 0}) \neq 0$.

\item Colour-spin adjoint, orbital singlet states
\begin{eqnarray}
&&\psi_{[Ad]}^{[S]}({\bf A}) = \lambda^a \phi_a ({\bf A}) \label{coloro} \\
&&\phi_a ({\bf A}^V) = \phi_a ({\bf A}) \, , \nonumber
\end{eqnarray}
with $\phi_a ({\bf 0}) \neq 0$ for some values of $a$. 

\item Colour-spin singlet, orbital $[\alpha]$ states with singlet source indices, and belonging to the irreducible colour representation $[\alpha]$ (see eq.~(\ref{alpha}))
\begin{eqnarray}
&&\psi^{[\alpha]}_{m\,[S]}({\bf A}) = {\phi^{[\alpha]}}_{m} ({\bf A}) I \label{wfuncany} \\
&& {\phi^{[\alpha]}}_{m}({\bf A}^V) = R^{[\alpha]}_{mm'} (V) {\phi^{[\alpha]}}_{m'} ({\bf A})
\, , \nonumber
\end{eqnarray}
with ${\phi^{[\alpha]}}_{m} ({\bf 0}) = 0$.

\item Colour-spin adjoint, orbital $[\beta]$ states giving rise to an irreducible colour representation $[\alpha]\in [\beta] \otimes [N_c^2-1]$ (see eq.~(\ref{beta}))
\begin{eqnarray}
&&\psi^{[\alpha]}_{m\,[Ad]}({\bf A}) = \lambda^a \phi_{a k} ({\bf A}) \label{wfuncany1} \\
&& {\phi_a}_k ({\bf A}^V) = R^{[\beta]}_{kk'} (V) \phi_{a k'} ({\bf A}) 
\, , \nonumber
\end{eqnarray}
with $\phi_{a k} ({\bf 0}) =0$. We repeat that in eq.~(\ref{wfuncany1}) $\phi_{a k} ({\bf A})$ is a tensor in the $[\beta] \otimes [N_c^2-1]$ space living in the $[\alpha]$-invariant subspace with values of $a$'s and $k$'s constrained so as to yield the desired value of the index $m$.
\end{itemize}
We remark that, since ${\bf A}^V = {\bf A}$ for $V\in Z_{N_c}$, where $Z_{N_c}$ is the center of SU$(N_c)$, the representation $[\alpha]$ in eqs.~(\ref{wfuncany}) and~(\ref{wfuncany1}) is actually bound to be a representation of SU$(N_c)/{Z_{N_c}}$.

The above classification is confirmed by perturbation theory~\cite{Rossi:1980pg}~\cite{Leroy:1982rg}~\cite{Leroy:1990eh}. States like those in eqs.~(\ref{colors}) and~(\ref{coloro}) correspond, at zero order in the coupling constant $g$, to wave functionals which do not contain any gluon component, and are therefore non vanishing at ${\bf A} = {\bf 0}$. Excited states consist of states with gluons added to external sources. The presence of gluons allows the energy levels to reach any zero $N_c$-ality (triality for $N_c=3$) representation. States of this kind, however, contain some power of the gluon fields, so that their wave functional vanishes at ${\bf A} = {\bf 0}$. As a working hypothesis, to be checked by non-perturbative numerical simulations, we conjecture that similar structures also describe the non-perturbative dynamics of the external source sectors. 

\subsection{The structure of the ${\bf A}$-traced propagation kernel} 
\label{sec:propstates}

In this section we determine the way in which the four classes of colour multiplets, discussed in sect.~\ref{sec:classst}, contribute to the Feynman propagation kernel traced over the field boundary values
\beqn
\hspace{-1.5cm}&&K_{s_2r_2;s_1r_1}(T) \equiv\int {\cal D}\mu_F({\bf A}) K({\bf A},s_2,r_2;{\bf A},s_1,r_1;T) \, . \label{DECOMP}
\eeqn
We remark that the gauge fixing in the ${\bf A}$ integration in eq.~(\ref{DECOMP}) is needed in order to make all the matrix elements of $K_{s_2r_2;s_1r_1}(T)$ finite. In fact, although $K({\bf A},s_2,r_2;{\bf A},s_1,r_1;T)$ is invariant only under gauge transformations which are equal to the identity at the location of the colour sources, the colour-traced kernel $\sum_{s r} K({\bf A},s,r;{\bf A},s,r;T)$ is invariant under {\em all} the time independent gauge transformations belonging to ${\cal G}_0$, and therefore $\sum_{sr}K_{sr;sr}(T)$ would be infinite in the absence of gauge fixing.

The contribution of the states given in eqs.~(\ref{colors}), (\ref{coloro}), (\ref{wfuncany}) and~(\ref{wfuncany1}) to the partial trace~(\ref{DECOMP}), can be computed from the expressions of the partial traces of the corresponding projectors given in Appendix~\ref{sec:APPB}. We find
\begin{itemize}
\item
Colour-spin singlet, orbital singlet states of the form~(\ref{colors})
\begin{eqnarray}
K_{s_2r_2;s_1r_1}(T) \Big{|}_{[S]}^{[S]}= \frac{\delta_{s_2 r_2}\delta_{r_1s_1}}{N_c} e^{-E^{[S]}T} \, .\label{colstrucsing} 
\end{eqnarray}
\item
Colour-spin adjoint, orbital singlet states of the form~(\ref{coloro}) 
\begin{eqnarray}
K_{s_2r_2;s_1r_1}(T)\Big{|}_{[Ad]}^{[S]}=2\sum_c {\lambda^c}_{s_2r_2} \lambda^c_{r_1s_1} \, e^{-E^{[Ad]} T} \, .\label{colstrucoct}
\end{eqnarray}
\item
Colour-spin singlet, orbital $[\alpha]$ states of the form~(\ref{wfuncany}) 
\begin{eqnarray}
K_{s_2r_2;s_1r_1}(T)\Big{|}^{[\alpha]}_{[S]}= \frac {D_{[\alpha]}}{N_c}\, \delta_{s_2 r_2} \delta_{r_1 s_1} e^{-E^{[\alpha]} T} \, .\label{colstrucsingalpha}
\end{eqnarray}
\item
Colour-spin adjoint, orbital $[\beta]$ states of the form~(\ref{wfuncany1}) with $[\alpha]\in [\beta] \otimes [N_c^2-1]$
\begin{eqnarray}
K_{s_2r_2;s_1r_1}(T)\Big{|}^{[\alpha]}_{[Ad]}= 2 \frac{D_{[\alpha]}}{N_c^2-1}\, \sum_c {\lambda^c}_{s_2 r_2} {\lambda^c}_{r_1 s_1} e^{-E^{[\alpha]} T} \, .\label{colstrucsingalphap}
\end{eqnarray}
\end{itemize}

\subsection{The structure of the $K({\bf 0},s_2,r_2;{\bf 0},s_1,r_1;T)$ kernel} 
\label{sec:propstatesz}

The previous considerations immediately imply that only the states in eqs.~(\ref{colors}) and~(\ref{coloro}) contribute to $K({\bf 0},s_2,r_2;{\bf 0},s_1,r_1;T)$, yielding terms with a tensor structure proportional to that of eqs.~(\ref{colstrucsing}) and~(\ref{colstrucoct}), respectively. In formulae we get (see eqs.~(\ref{prjsingst}) and~(\ref{PAD}))
\beqn
\hspace{-.8cm}&&K({\bf 0},s_2,r_2;{\bf 0},s_1,r_1;T) = \nonumber \\
\hspace{-.8cm}&&=|\phi({\bf 0})|^2\,\frac{\delta_{s_2 r_2}\delta_{r_1s_1}}{N_c} \,e^{-E^{[S]}T} + \sum_{a} |\phi_a ({\bf 0})|^2 \sum_b\lambda^b_{s_2r_2}\lambda^b_{r_1s_1} e^{-E^{[Ad]} T} + \ldots\, .
\label{KZZ}
\eeqn
From the structure of states summarized in equations from~(\ref{colors}) to~(\ref{wfuncany1}) and the results in perturbation theory~\cite{Rossi:1980pg}~\cite{Leroy:1982rg}~\cite{Leroy:1990eh}, we conjecture that only the first two terms displayed in eq.~(\ref{KZZ}) are actually present. This statement can be tested in lattice numerical simulations. 

If we were only interested in the computation of the $q\bar q$-potential in the vacuum, the relevant quantity to study would be $K({\bf 0},s_2,r_2;{\bf 0},s_1,r_1;T)$. Instead, in order to explore more complicated situations such as, for instance, the $q\bar q$-potential at finite temperature, the relevant quantity to consider is $K_{s_2r_2;s_1r_1}(T)$. 

\section{Extracting singlet and adjoint potential in the continuum}
\label{sec:SINOCT}

To extract the lowest energy eigenvalues of the singlet and adjoint channel we have to project eq.~(\ref{DECOMP}) (or eq.~(\ref{KZZ})) over the $\delta_{r_2s_2} \delta_{s_1r_1}/N_c$ and $2\sum_a \lambda^a_{r_2s_2}\lambda^a_{s_1r_1}$ tensor projector (see Appendix~\ref{sec:APPC}), respectively. 

Referring to the partially traced kernel~(\ref{DECOMP}) and using eqs.~(\ref{colstrucsing}) and~(\ref{colstrucsingalpha}), we obtain at large times for the ``colour-spin singlet"  channels (see eqs.~(\ref{colors}) and~(\ref{wfuncany})) 
\beqn
\hspace{-1.5cm}&&\sum_{s_2r_2s_1r_1}\frac{1}{N_c}\delta_{r_2s_2}\delta_{s_1r_1}K_{s_2r_2;s_1r_1}(T) \stackrel{T\to \infty}\rightarrow 
e^{-E^{[S]}T}+\ldots +D_{[\alpha]}e^{-E^{[\alpha]}T}+\ldots\, ,
\label{ES}
\eeqn
where the dots stand for exponentially suppressed terms and, as discussed before, $D_{[\alpha]}$ is the dimension of the colour representation $[\alpha]$ to which the energy eigenstate belongs. 

Similarly, using eqs.~(\ref{colstrucoct}) and~(\ref{colstrucsingalphap}), for the ``colour-spin adjoint" channels (see eqs.~(\ref{coloro}) and~(\ref{wfuncany1})) we find
\beqn
\hspace{-1.cm}&&\sum_{s_2r_2s_1r_1}2\sum_a \lambda^a_{r_2s_2}\lambda^a_{s_1r_1}K_{s_2r_2;s_1r_1}(T) \stackrel{T\to \infty}\rightarrow 
(N_c^2-1)\,e^{-E^{[Ad]}T}+\ldots +D_{[\alpha']}e^{-E^{[\alpha']}T}+\ldots\, .
\label{EAD}
\eeqn
In perturbation theory the ``singlet and adjoint $q\bar q$-potentials" defined in eqs~(\ref{ES}) and~(\ref{EAD}) contain divergent self-energy contributions, independent of  the relative position of the sources. In perturbation theory these contributions can be renormalized away dividing the Feynman kernel~(\ref{DECOMP}) by the product of the two traces (characters) of the kernels with the insertion of a single $q$ or $\bar q$ source. This can be done by computing, at large $T$, the quantity 
\beqn
\hspace{-.8cm}&&\Gamma_{s_2r_2;s_1r_1}(T) = \frac{K_{s_2r_2;s_1r_1}(T)}{K_q(T)K_{\bar q}(T)}\, ,\label{RA}
\eeqn
where
\beqn
\hspace{-.8cm}&& K_q(T) = \int_{{\cal G}_0}{\cal D}\mu(h) \tr \big{[}e^{i\lambda^a h^a({\bf x}_q)}\big{]}\int {\cal D}\mu_F({\bf A}) \widetilde K({\bf A}^{U_h},{\bf A};T) 
\, ,\label{SQ}\\
\hspace{-.8cm}&& K_{\bar q}(T) = \int_{{\cal G}_0}{\cal D}\mu(h) \tr\big{[}e^{-i\lambda^a h^a({\bf x}_{\bar q})}\big{]}\int {\cal D}\mu_F({\bf A}) \widetilde K({\bf A}^{U_h},{\bf A};T) \, .\label{SQB}
\eeqn
We stress again that, had we included in the gauge average also the integration over global colour rotations, the states in the spectral decomposition of the averaged Feynman kernel corresponding to colour non-singlet states would be missing (see eq.~(\ref{INVACS})). Consequently, in the large $T$ limit, we might not be anymore in position to reach the desired lowest energy eigenvalues. Another unwanted consequence is that colour averaging makes the Feynman kernel contribution of a state vanish when the tensor product of source representations does not contain the singlet. This is precisely what would happen to the factors in the denominator of eq.~(\ref{RA}). 

In the non-perturbative regime, where colour is supposed to be confined, the quantities defined in eqs.~(\ref{SQ}) and~(\ref{SQB}) could be zero anyway. So the need and the way in which this particular renormalization step should be carried out must be the object of numerical investigations.

\section{Extracting the $q\bar q$-potential from lattice simulations}  
\label{sec:LTS} 

In this section we discuss how to compute, in lattice simulations, the singlet and adjoint $q\bar q$-potentials. Although it is not difficult to translate in lattice language the temporal gauge fixing procedure presented in the previous sections and provide a discretized version of eqs.~(\ref{DECOMP}) or~(\ref{KZZ}), it turns out that these formulae are not well suited to the structure of practical numerical simulations. 

The reason is that it is not easy to implement on the lattice the condition $U_h\in {\cal G}_0$ (i.e.\ $U_h({\bf x})\stackrel {{\bf |x|}\to \infty} \longrightarrow I$) on the boundary gauge integration in eq.~(\ref{KQQB}). The limitation $U_h\in {\cal G}_0$ is, however, absolutely crucial, as we have discussed, in order to avoid the cancellation from the colour averaged kernel of all the global colour non-singlet states (if not dynamically confined).  

The way out to this difficulty is to isolate in the Feynman propagation kernel on the lattice the contribution of states belonging to given representations, averaging with the SU($N_c$) characters. This idea is based on the formula
\beqn
&&\sum_{sr}\int {\cal D}\mu_F({\bf A})K({\bf A}^V, s,r;{\bf A}, s,r;T)=\nonumber\\
&&=\sum_{sr}\int {\cal D}\mu_F({\bf A})\sum_{k,m} \psi^{[\alpha_k]}_m({\bf A}^V,s,r) (\psi^{[\alpha_k]}_m({\bf A},s,r))^*e^{-E^{[\alpha_k]}T}=\nonumber\\
&&=\sum_{sr}\int {\cal D}\mu_F({\bf A})\sum_{k,m,m'} R_{m m'}^{[\alpha_k]}(V)\psi^{[\alpha_k]}_{m'}({\bf A},s,r) (\psi^{[\alpha_k]}_m({\bf A},s,r))^*e^{-E^{[\alpha_k]}T}=\nonumber\\
&&=\sum_k \tr[R^{[\alpha_k]}(V)]e^{-E^{[\alpha_k]}T}\equiv \sum_k \chi^{[\alpha_k]}(V)]e^{-E^{[\alpha_k]}T}\, ,
\label{KEXPSPE}
\eeqn
which shows that performing a global colour rotation on one of the boundary gauge field and taking the full trace of the kernel amounts to replacing the multiplicity factor appearing in eq.~(\ref{tottrace}) with the character of the irreducible representation to which each energy eigenfunctional belongs. 

\subsection{Temporal lattice gauge fixing}  
\label{sec:TGF} 

We start discussing how the temporal gauge fixing procedure can be implemented on the lattice. Link variables are generated by some Monte Carlo algorithm with weight given by (minus) the exponent of the (gauge unfixed) plaquette action, assuming periodic boundary conditions in the four dimensions. Lattice points are denoted by $\{n_0,{\bf n}\}$, with $n_0=0,1,\ldots,N_T$, $n_k=1,\ldots,N_L$, $k=1,2,3$.

On each configuration the temporal gauge fixing is carried out subjecting the gauge links to the gauge transformation
\beqn
\hspace{-1.cm}&&\Omega({\bf n},t)= U^\dagger_0({\bf n},t-1;{\bf n},t)\ldots
U_0^\dagger({\bf n},1;{\bf n},2) U_0^\dagger({\bf n},0;{\bf n},1)\, , \quad t=1,\ldots,N_T \, .
\label{GTTZZ} \\
\hspace{-1.cm}&&\Omega({\bf n},0)=I \, ,
\label{GTTZ} 
\eeqn
which sets $U_0({\bf n},0;{\bf n},t)=I$ at $t=1,\ldots,N_T-1$. Under the gauge transformation~(\ref{GTTZZ}) the spatial links transform as 
\beqn
\hspace{-1.5cm}&&U_k({\bf n},t;{\bf n}+\hat {\bf k},t)\,\to \,U'_k({\bf n},t;{\bf n}+\hat {\bf k},t)=\Omega^\dagger({\bf n},t)U_k({\bf n},t;{\bf n}+\hat {\bf k},t)\Omega({\bf n}+ \hat {\bf k},t)\, , \label{GTSZ} \\
\hspace{-1.5cm}&&\qquad\qquad\qquad\qquad\quad t=1,\ldots,N_T \, .\nonumber
\eeqn
Notice that, owing to eq.~(\ref{GTTZ}), the spatial links at $t=0$ are untouched. 

The time ordered product of the temporal links (open Polyakov lines) that represent (in the unfixed theory, see also eq.~(\ref{KQQBC}) of Appendix~\ref{sec:APPD}) the external $q$ and $\bar q$ colour sources at ${\bf n}_q$ and ${\bf n}_{\bar q}$, become
\beqn
\hspace{-1.0cm}&&\prod_{t=1}^{N_T}U_0({\bf n}_q,t-1;{\bf n}_q,t)=\nonumber\\
\hspace{-1.0cm}&&\quad=\prod_{t=1}^{N_T}\Omega({\bf n}_q,t-1)U'_0({\bf n}_q,t-1;{\bf n}_q,t)\Omega^\dagger({\bf n}_q,t)=\Omega^\dagger({\bf n}_q,N_T)\equiv \Omega^\dagger({\bf n}_{q}) \, ,\label{PLQ}\\
\hspace{-1.0cm}&&\Big{[}\prod_{t=1}^{N_T}U_0({\bf n}_{\bar q},t-1;{\bf n}_{\bar q},t)\Big{]}^\dagger=\nonumber\\
\hspace{-1.0cm}&&\quad=\Big{[}\prod_{t=1}^{N_T}\Omega({\bf n}_{\bar q},t-1)U'_0({\bf n}_{\bar q},t-1;{\bf n}_{\bar q},t)\Omega^\dagger({\bf n}_{\bar q},t)\Big{]}^\dagger=\Omega({\bf n}_{\bar q},N_T)\equiv \Omega({\bf n}_{\bar q})  \, .\label{PLQB}
\eeqn
Configuration averages performed at this stage will inevitably also include an average over global colour rotations. In the next subsection we describe a method based on character projections which exploits this colour average to select the contribution to the Feynman propagation kernel coming from states belonging to a given global colour representation.

\subsection{Character projection}  
\label{sec:CHA} 

As discussed above, we can easily compute on the lattice the colour averaged kernel (see eq.~(\ref{KQQBAD})), which in continuum notation reads 
\begin{eqnarray}
&&\overline K_{s_2,r_2;s_1,r_1}(T) \equiv \int {\cal D} \mu_F({\bf A}) \int_{\overline{\cal G}_0}{\cal D}\mu(h)  \Big{[}e^{i\lambda^a h^a({\bf x}_q)}\Big{]}_{s_2s_1}\Big{[}e^{i\lambda^a h^a({\bf x}_{\bar q})}\Big{]}^*_{r_2r_1}\widetilde K({\bf A_2}^{U_h},{\bf A_1};T)\, .\label{KAVB}\end{eqnarray}
Within the $\overline{\cal G}_0$-integration we can factorize the integration over ${\cal G}_0$ times the integration over the SU$(N_c)$ group, getting
\begin{eqnarray}
&&\overline K_{s_2,r_2;s_1,r_1}(T) =\int_{{\rm SU}(N_c)} {\cal D} V\, V_{s_2s_3}V_{r_2r_3}^*  \int {\cal D} \mu_F({\bf A}) K({\bf A}^V,s_3,r_3;{\bf A},s_1,r_1;T)  \, ,\label{KAV}
\end{eqnarray}
where $K({\bf A},s_3,r_3;{\bf A},s_1,r_1;T)$ is the original Feynman kernel in the temporal gauge defined in eq.~(\ref{KQQBAD}).

We will now show that the information about $q\bar q$-potentials can be extracted computing the character-weighted kernel  
\begin{eqnarray}
\hspace{-.8cm}{\overline K^{[\gamma]}}_{s_2,r_2;s_1,r_1}(T) \equiv\int_{{\rm {SU}}(N_c)} \!\!{\cal D} V \,(\chi^{[\gamma]}(V))^*V_{s_2s_3}V_{r_2r_3}^*  \int {\cal D} \mu_F({\bf A}) K({\bf A}^V,s_3,r_3;{\bf A},s_1,r_1;T) \, ,\label{CHARK}
\end{eqnarray}
where $\chi^{[\gamma]}(V)$ is the character of the representation $[\gamma]$. 
In the following we will also be interested in the total trace
\begin{eqnarray}
{\overline K^{[\gamma]}}(T) \equiv \sum_{rs} {\overline K^{[\gamma]}}_{s,r;s,r}(T)\, ,\label{TT}
\end{eqnarray}
for which, in virtue of its ${\cal G}_0$ invariance, we have the further benefit that there is no need for a lattice gauge fixing while computing the boundary ${\bf A}$-integration.

The proof of these facts is based on the orthogonality property of characters 
\beq
\int_{{\rm{SU}}(N_c)} \!\!{\cal D} V \,(\chi^{[\alpha]}(V))^* \chi^{[\beta]}(V) =\delta_{[\alpha],[\beta]}\, ,
\label{CHARA}
\eeq
where $\chi^{[\alpha]}(V)$ and $\chi^{[\beta]}(V)$ are the characters of the irreducible representations $[\alpha]$ and $[\beta]$.

To proceed we need to study the contribution to ${K^{[\gamma]}}_{s_2,r_2;s_1,r_1}(T)$ and its total trace of the four types of energy eigenstates~(\ref{colors})--(\ref{wfuncany1}) discussed in sect.~\ref{sec:EXC}. 

\begin{itemize}

\item From~(\ref{colors}) the colour-spin singlet, orbital singlet state contribution to ${\overline K^{[\gamma]}}_{s_2,r_2;s_1,r_1}(T)$ is seen to be
\begin{eqnarray}
{\overline K^{[\gamma]}}_{s_2,r_2;s_1,r_1}(T)\Big{|}_{[S]}^{[S]} = \delta_{[\gamma],[S]} \frac{\delta_{s_2 r_2} \delta_{r_1s_1}}{N_c} e^{-E^{[S]}T} \, .\label{KSS}
\end{eqnarray}
Thus for the fully traced kernel we get 
\begin{eqnarray}
{\overline K^{[\gamma]}}(T)\Big{|}_{[S]}^{[S]} = \delta_{[\gamma],[S]} e^{-E^{[S]}T}\, .\label{FTKSS}
\end{eqnarray}

\item From~(\ref{partialtr8}) we find that the contribution of colour-spin adjoint, orbital singlet states~(\ref{coloro}) to the character-weighted kernel is 
\begin{eqnarray}
&&{\overline K^{[\gamma]}}_{s_2,r_2;s_1,r_1}(T)\Big{|}^{[S]}_{[Ad]}  = 2 \int_{{\rm {SU}}(N_c)} {\cal D} V (\chi^{[\gamma]}(V))^* \sum_c (V \lambda^c V^\dagger)_{s_2 r_2} (\lambda^c_{s_1 r_1})^* e^{-E^{[Ad]}T} = \nonumber \\
&&= 2\int_{{\rm{SU}}(N_c)} {\cal D} V (\chi^{[\gamma]}(V))^* \sum_{a c} R^{[N_c^2-1]}_{ac} (V) \lambda^a_{s_2 r_2} (\lambda^c_{s_1 r_1})^* e^{-E^{[Ad]}T} \, . \label {octsingl}
\end{eqnarray}
The r.h.s.\ of eq.~(\ref{octsingl}) is different from zero only if $[\gamma] = [N_c^2-1]$. Tracing with $\delta_{s_2s_1}\delta_{r_2r_1}$ the resulting trace over the $\lambda$ matrices provides the factor $\delta_{ac}/2$ that brings in the character of the adjoint representation $\sum_a R^{[N_c^2-1]}_{aa} (V)=\chi^{[N_c^2-1]}(V)$. From the orthogonality of inequivalent characters we thus get for the fully traced kernel 
\begin{eqnarray}
\hspace{-.5cm}{\overline K^{[\gamma]}}(T) \Big{|}^{[S]}_{[Ad]} = \int_{{\rm{SU}}(N_c)} {\cal D} V (\chi^{[\gamma]}(V))^* \chi^{[N_c^2-1]} (V) e^{-E^{[Ad]}T} = \delta_{[\gamma],[N_c^2-1]} e^{-E^{[Ad]}T}\, .\label{FTKSSAD}
\end{eqnarray}

\item Recalling the form of the projector~(\ref{PRO2}), we see that the contribution of the  colour-spin singlet, orbital $[\alpha]$ states~(\ref{wfuncany}) to the character-weighted kernel is 
\begin{eqnarray}
&&{\overline K^{[\gamma]}}_{s_2,r_2;s_1,r_1}(T)\Big{|}^{[\alpha]}_{[S]}  = \nonumber\\
&& = \delta_{s_2 r_2} \delta_{r_1 s_1} \int_{{\rm {SU}}(N_c)} \!{\cal D} V (\chi^{[\gamma]}(V))^* \sum_m \int {\cal D} \mu_F({\bf A}) {\phi^{[\alpha]}}_m({\bf A}^V) {\phi^{[\alpha]}}_m({\bf A})^* e^{-E^{[\alpha]}T} = \nonumber \\
&&= \delta_{s_2 r_2} \delta_{r_1 s_1} \int_{{\rm {SU}}(N_c)} \!{\cal D} V (\chi^{[\gamma]}(V))^* \sum_{mm'} R^{[\alpha]}_{mm'} (V)\int {\cal D} \mu_F({\bf A}) {\phi^{[\alpha]}}_{m'}({\bf A}) {\phi^{[\alpha]}}_m({\bf A})^* e^{-E^{[\alpha]}T} = \nonumber \\
&&= \frac{\delta_{s_2 r_2} \delta_{r_1 s_1}}{N_c} \int_{{\rm {SU}}(N_c)} \!{\cal D} V (\chi^{[\gamma]}(V))^* \chi^{[\alpha]} (V) = \frac{\delta_{s_2 r_2} \delta_{r_1 s_1}}{N_c} \delta_{[\gamma],[\alpha]}e^{-E^{[\alpha]}T}\, , \label{KAGA}
\end{eqnarray}
where we have used the orthogonality of the wave functionals ${\phi^{[\alpha]}}_m({\bf A})$ for different Cartan indices. For the fully traced kernel we get 
\begin{eqnarray}
{\overline K^{[\gamma]}}(T) \Big{|}^{[\alpha]}_{[S]} = \delta_{[\gamma],[\alpha]} e^{-E^{[\alpha]}T} \, .\label{KAGAT}
\end{eqnarray}

\item Finally using eq.~(\ref{PRO4}) and the orthogonality of $\phi_{am}({\bf A})$ wave functionals for different Cartan indices, we find that the contribution of the colour-spin adjoint, orbital $[\beta]$ states (eq.~(\ref{wfuncany1})), composing into the irreducible colour representation $[\alpha]$ to the character-weighted kernel is
\begin{eqnarray}
&&{\overline K^{[\gamma]}}_{s_2,r_2;s_1,r_1}(T)\Big{|}^{[\alpha]}_{[Ad]} = \nonumber\\
&&= \int_{{\rm SU}{(N_c)}} \!\!{\cal D} V (\chi^{[\gamma]}(V))^* \!\!\!\sum_{a k k'bc}\! \!{R_{bk;ck'}^{[\alpha]}}(V) \!\int \!{\cal D} \mu_F({\bf A}) \phi_{a k'}({\bf A}) \phi_{a k}({\bf A})^* \lambda^c_{s_2 r_2} ({\lambda^b}_{s_1 r_1})^* e^{-E^{[\alpha]}T} =\nonumber \\
&&= 2 \int_{{\rm SU}{(N_c)}} \!{\cal D} V (\chi^{[\gamma]}(V))^* \sum_{kbc} {R_{bk;ck}^{[\alpha]}}(V) \lambda^c_{s_2 r_2} ({\lambda^b}_{s_1 r_1})^* e^{-E^{[\alpha]}T} \label{KAGG}
\end{eqnarray}
and is different from zero only if $[\gamma] = [\alpha]$. For the fully traced kernel we get 
\begin{eqnarray}
&&{\overline K^{[\gamma]}}(T) \Big{|}^{[\alpha]}_{[Ad]}= \int_{{\rm SU}{(N_c)}} {\cal D} V (\chi^{[\gamma]}(V))^* \sum_{bk} {R_{bk;bk}^{[\alpha]}}(V) e^{-E^{[\alpha]}T} =\nonumber\\
&&= \int_{{\rm SU}{(N_c)}} {\cal D} V (\chi^{[\gamma]}(V))^* \chi^{[\alpha]}(V) e^{-E^{[\alpha]}T} =\delta_{[\gamma],[\alpha]} e^{-E^{[\alpha]}T}  \, , \label{KKBA}
\end{eqnarray}
since $\sum_{bk}{R_{bk;bk}^{[\alpha]}}(V)=\chi^{[\alpha]}(V)$ is the character of the representation $[\alpha]$.
\end{itemize}

\subsection{Extracting $q\bar q$-potential energies from lattice data}  
\label{sec:EELD} 

To extract the interesting $q\bar q$-potential energies from lattice data we propose to use the lattice version of (the large $T$-limit of) eqs.~(\ref{FTKSS}) and~(\ref{FTKSSAD}) for the singlet and adjoint representation, respectively. These formulae simply read
\beqn
\hspace{-.9cm}&&\bullet \,\overline K^{\chi^{[S]}}(T)\Big{|}_{[S]\,{\rm lat}}^{[S]}\!=\frac{1}{N_{conf}}\!\sum_{\ell=1}^{N_{conf}} \Big{[}\tr[\Omega^\dagger({\bf n}_q)]\tr[\Omega({\bf n}_{\bar q})]\Big{]}^{(\ell)}\!+\ldots \stackrel{T\to \infty}\rightarrow e^{-E^{[S]}T}+\ldots\, ,\label{AVERS}\\
\hspace{-.9cm}&&\bullet\, \overline K^{\chi^{[Ad]}}(T)\Big{|}_{[S]\,{\rm lat}}^{[Ad]}\!=\nonumber\\
&&\quad=\frac{1}{N_{conf}}\!\sum_{\ell=1}^{N_{conf}} (\chi^{[Ad]}(\Omega({\bf n}_\infty))^*
\Big{[}\tr[\Omega^\dagger({\bf n}_q)]\tr[\Omega({\bf n}_{\bar q})]\Big{]}^{(\ell)}\!+\ldots\stackrel{T\to \infty}\rightarrow e^{-E^{[Ad]}T}+\ldots\, ,\label{AVERAD}
\eeqn
where $N_{conf}\, (\gg 1)$ is the number of gauge configurations that have been generated,
and the dots represent corrections due to the finiteness of $N_{conf}$ as well as terms exponentially suppressed in the large $T$ limit. In eq.~(\ref{AVERAD}) ${\bf n}_\infty$ denotes the lattice point at infinity, far from the location of the colour sources, and $\Omega({\bf n}_\infty)$ identifies the global colour integration, unavoidable in lattice simulations, mentioned at the end of sect.~\ref{sec:GLOB} (see also the step from eq.~(\ref{KAVB}) to~(\ref{KAV}) where $V$ is what we are calling here $\Omega({\bf n}_\infty)$).

Naturally the character of the $[S]$ representation is unit, while for the adjoint character we have $\chi^{[Ad]}(V)=2 \sum_a\tr[\lambda^aV\lambda^aV^\dagger]$.

For the $q$ and ${\bar q}$ self-energy we find
\beqn
\hspace{-.9cm}&&\overline K^{[q]}(T)\Big{|}_{{\rm lat}}\!=\frac{1}{N_{conf}}\!\sum_{\ell=1}^{N_{conf}} (\chi^{[q]}(\Omega({\bf n}_\infty))^* \tr\Big{[}\Omega^\dagger({\bf n}_q)\Big{]}^{(\ell)}\!+\ldots \stackrel{T\to \infty}\rightarrow e^{-E^{[q]}T}+\ldots\, ,\label{AVERQS}\\
\hspace{-.9cm}&&\overline K^{[\bar q]}(T)\Big{|}_{{\rm lat}}\!=\frac{1}{N_{conf}}\!\sum_{\ell=1}^{N_{conf}} (\chi^{[\bar q]}(\Omega({\bf n}_\infty))^*\tr\Big{[}\Omega({\bf n}_{\bar q})\Big{]}^{(\ell)}\!+\ldots\stackrel{T\to \infty}\rightarrow e^{-E^{[\bar q]}T}+\ldots\, ,\label{AVERAQD}
\eeqn
where, we recall, $\chi^{[q]}(V)=\tr[V]$ and $\chi^{[\bar q]}(V)=\tr[V^\dagger]$. 

\section{Conclusions and outlook}
\label{sec:CONCL}

In this paper we have derived explicit expressions for the singlet and the octet (adjoint)
potential between two static, point-like sources with the colour quantum numbers of a $q\bar q$-pair (see eqs.~(\ref{AVERS}) and~(\ref{AVERAD})). They have a particularly transparent form in the temporal gauge. For completeness in Appendix~\ref{sec:APPD} we provide the formulae valid in the Coulomb gauge. 

We have discussed in sect.~\ref{sec:GLOB} the importance of limiting the gauge integration in eq.~(\ref{KQQB}) to the gauge transformations that tend to the identity at spatial infinity, showing that otherwise only global colour singlet eigenfunctionals would contribute to the  colour averaged Feynman kernel. 

In sect.~\ref{sec:LTS} we have shown how the temporal gauge fixing can be implemented in practical lattice simulations and how it is possible to extract the singlet and adjoint $q\bar q$-potential energy from lattice simulations, weighting the fully traced Feynman kernel with the character of the representation one is interested in filtering out.

Explicit numerical lattice simulations are under way~\cite{PREP} to check to validity of the analysis presented in this paper and the viability of the formulae we have derived for extracting singlet and octet $q\bar q$-potential. 

\vspace{.4cm}
{\bf Acknowledgements - }We thank K.\ Jansen for bringing the problem addressed in this paper to our attention and for discussions. We acknowledge MIUR (Italy) for partial support under the PRIN contract number 20093BMNPR. One of us (GCR) wish to thank the Galileo Galilei Institute for Theoretical Physics for the hospitality during the workshop "New Frontiers in Lattice Gauge Theories", while this work was initiated.

\appendix 

\section{Completeness and Group Integration}  
\label{sec:APPA}

When the degeneracy of an energy level is due to the existence of a symmetry group, it is possible to describe the completeness sum over the degeneracy subspace in terms of the invariant group integration. The basic idea is easily explained as follows.

Starting from an irreducible representation $R(g)$ of a group ${G}$ and a basis $|i\rangle$ of the representation space, we have the definition 
\begin{eqnarray}
R(g) |i \rangle = \sum_j |j\rangle\langle j| R(g) |i\rangle = \sum_j R_{j i}(g) |j\rangle \, .
\end{eqnarray}
The completeness relation can be expressed as
\begin{eqnarray}
I = \sum_i |i\rangle\langle i| = D_{[R]} \int_G {\cal D}g R(g) |{\bar k} \rangle\langle {\bar k}| R^\dagger (g) \, , \label{DIMENSION}
\end{eqnarray}
with $|{\bar k}\rangle$ an arbitrary representation state. In eq.~(\ref{DIMENSION}) $D_{[R]}$ is the dimension of the space spanned by the representation $[R]$. In fact, from (see ref.~\cite{LIU})
\begin{eqnarray}
\int_G{\cal D}g R_{j k}(g) R_{l m} (g)^* = \frac{\delta_{jl} \delta_{km}}{D_{[R]}} \, , \label{pweyl}
\end{eqnarray}
we get
\begin{eqnarray}
\hspace{-.8cm}D_{[R]} \int_G{\cal D}g R(g) |{\bar k} \rangle\langle {\bar k}| R^\dagger (g) = 
D_{[R]} \sum_{j,l} \int_G {\cal D}g R_{j {\bar k}}(g) | j \rangle\langle l | R_{l {\bar k}} (g)^* = \sum_{l} | l \rangle\langle l | \, .
\end{eqnarray}
The interest of eq.~(\ref{DIMENSION}) is that it allows to write the completeness relation by only knowing a single (arbitrary) representation state, $|{\bar k} \rangle$.

\section{Colour Projectors} 
\label{sec:APPB}

We apply the formalism developed in Appendix~\ref{sec:APPA} to write the colour projectors necessary to single out the states listed in eqs.~(\ref{colors}), (\ref{coloro}), (\ref{wfuncany}) and~(\ref{wfuncany1}).

$\bullet$ We start with the normalization condition for the colour-spin singlet, orbital singlet states of eq.~(\ref{colors}) which reads 
\begin{eqnarray}
1=  \int {\cal D} \mu_F ({\bf A}) \tr \Big{[}|\psi^{[S]}_{[S]}({\bf A})^\dagger \psi^{[S]}_{[S]}({\bf A})\Big{]} = 
N_c  \int {\cal D} \mu_F ({\bf A}) |\phi ({\bf A})|^2 \, . \label{normsing}
\end{eqnarray}
The projector over a colour-spin singlet, orbital singlet state is given by
\begin{eqnarray}
 {\cal P}^{[S]}_{[S]}=\phi ({\bf A}_2) \phi^* ({\bf A}_1) \delta_{s_2 r_2}\delta_{r_1s_1}  \, , \label{prjsingst}
\end{eqnarray}
so that it contributes to the partially traced kernel~(\ref{DECOMP}) with a term
\begin{eqnarray}
\int {\cal D} \mu_F ({\bf A}) | \phi ({\bf A}) |^2 \,\delta_{s_2 r_2} \delta_{r_1s_1} e^{-E^{[S]}T}= \frac{\delta_{s_2 r_2} \delta_{r_1s_1}}{N_c} e^{-E^{[S]}T} \, , \label{partialtr0}
\end{eqnarray}
as it follows from eq.~(\ref{normsing}).

$\bullet$ The normalization condition for the colour-spin adjoint, orbital singlet states of eq.~(\ref{coloro}) is 
\begin{eqnarray}
&&1=  \int {\cal D} \mu_F ({\bf A}) \tr \Big{[}\psi_{[Ad]}^{[S]}({\bf A})^\dagger {\psi_{[Ad]}^{[S]}}({\bf A})\Big{]} = \nonumber \\
&& = \tr [\lambda^a \lambda^b] \int {\cal D} \mu_F ({\bf A}) \phi_a ({\bf A}) \phi_b^* ({\bf A}) = \frac{1}{2} \sum_a \int {\cal D} \mu_F ({\bf A}) |\phi_a ({\bf A})|^2 \, . \label{normoct}
\end{eqnarray}
The projector over the colour-spin adjoint, orbital singlet states can be computed with the help of the group integration formula~(\ref{pweyl}) obtaining (we recall that $R_{ab}^{[Ad]}(V)$ is a real matrix)

\begin{eqnarray}
&& {\cal P}_{[Ad]}^{[S]}=(N_c^2-1) \int_{{\rm SU}{(N_c)}}  {\cal D} V \,(V \psi_{[Ad]}^{[S]}({\bf A}_2^V) V^\dagger)_{s_2 r_2}  (V \, \psi_{[Ad]}^{[S]}({\bf A}_1^V) V^\dagger)^*_{s_1 r_1} = \nonumber \\
&&= (N_c^2-1) \int_{{\rm SU}{(N_c)}}{\cal D} V \,(V \lambda^a V^\dagger)_{s_2 r_2} (V \lambda^b V^\dagger)^*_{s_1 r_1} \phi_a ({\bf A}_2)\phi_b ({\bf A}_1)^* = \nonumber \\
&& = (N_c^2-1) \int_{{\rm SU}{(N_c)}}{\cal D} V \, R_{ca}^{[Ad]} (V) \lambda^c_{s_2 r_2} R_{db}^{[Ad]} (V) (\lambda^d_{s_1 r_1})^*  \phi_a ({\bf A}_2)\phi_b ({\bf A}_1)^*=\nonumber\\
&&= \sum_a\phi_a ({\bf A}_2)\phi_a({\bf A}_1)^* \sum_c \lambda^c_{s_2 r_2} (\lambda^c_{s_1 r_1})^*\, . \label{PAD}
\end{eqnarray}
Using the normalization condition~(\ref{normoct}), we conclude that a colour-spin adjoint, orbital singlet state contributes to the partially traced kernel of eq.~(\ref{KQQBV1}) with a term
\begin{eqnarray}
\sum_a \int {\cal D}\mu_F({\bf A}) |\phi_a ({\bf A})|^2 \sum_c \lambda^c_{s_2 r_2} (\lambda^c_{s_1 r_1})^* e^{-E^{[Ad]}T} = 2 \sum_c \lambda^c_{s_2 r_2} \lambda^c_{r_1 s_1} e^{-E^{[Ad]}T} \, . \label{partialtr8}
\end{eqnarray}
As a check of eqs.~(\ref{partialtr0}) and~(\ref{partialtr8}) we can trace them over colour source indices obtaining $1$ for the singlet and $(N_c^2-1)$ for the adjoint state, in agreement with the fact that the trace of a projector is the dimension of the space over which it projects.

$\bullet$ The normalization condition for the colour-spin singlet, orbital $[\alpha]$ states of eq~(\ref{wfuncany}) is (no sum over $m$)
\begin{eqnarray}
1 = \int {\cal D}\mu_F({\bf A}) \, \tr \Big{[}\psi^{[\alpha]}_{[S]m}({\bf A})^\dagger\psi^{[\alpha]}_{[S]m}({\bf A})\Big{]} = N_c \, \int {\cal D}\mu_F({\bf A}) | {\phi^{[\alpha]}}_m ({\bf A})|^2 \, .
\end{eqnarray}
Eq.~(\ref{DIMENSION}) gives for the projector over the $\psi^{[\alpha]}_{[S]m}$ multiplet the expression
\begin{eqnarray}
{\cal P}^{[\alpha]}_{[S]}=D_{[\alpha]} \int_{{\rm SU}{(N_c)}} {\cal D} V \, {\phi^{[\alpha]}}_m({\bf A}_2^V) {\phi^{[\alpha]}}_m({\bf A}_1^V)^* \delta_{s_2 r_2} \delta_{r_1 s_1} \label{PRO1}
\end{eqnarray}
for any fixed value of $m$. From the transformation properties~(\ref{wfuncany}) and using eq.~(\ref{pweyl}), we can further elaborate eq.~(\ref{PRO1}) with the result
\begin{eqnarray}
&&{\cal P}^{[\alpha]}_{[S]}=D_{[\alpha]} \int_{{\rm SU}{(N_c)}} {\cal D} V \, R^{[\alpha]}_{mm'}(V){\phi^{[\alpha]}}_{m'}({\bf A}_2) (R^{[\alpha]}_{mm''}(V){\phi^{[\alpha]}}_{m''}({\bf A}_1))^* \delta_{s_2 r_2} \delta_{r_1 s_1} =\nonumber\\
&&=\sum_m {\phi^{[\alpha]}}_{m}({\bf A}_2) {\phi^{[\alpha]}}_{m}({\bf A}_1)^*\delta_{s_2 r_2} \delta_{r_1 s_1}\, .\label{PRO2}
\end{eqnarray}
Starting with eq.~(\ref{PRO1}), we now compute the contribution of this state multiplet to the ${\bf A}$ partial trace of the kernel, finding 
\begin{eqnarray}
&&D_{[\alpha]} \int_{{\rm SU}{(N_c)}}  {\cal D} V \int {\cal D}\mu_F({\bf A}) \, {\phi^{[\alpha]}}_{m}({\bf A}^V) {\phi^{[\alpha]}}_m({\bf A}^V)^* \delta_{s_2 r_2} \delta_{r_1 s_1} e^{-E^{[\alpha]}T} = \nonumber \\ 
&&=\frac {D_{[\alpha]}}{N_c} \delta_{s_2 r_2} \delta_{r_1 s_1} e^{-E^{[\alpha]}T} \, .\label{DA}
\end{eqnarray}
To derive eq.~(\ref{DA}) we have exploited the colour invariance of the measure ${\cal D}\mu_F({\bf A})$ and the normalization of the group measure, $\int_{{\rm SU}{(N_c)}}{\cal D}V=1$. As before, the total trace gives the dimension of the representation space, $D_{[\alpha]}$.

$\bullet$ In order to apply the same procedure to the colour-spin adjoint, orbital $[\alpha]$ states of eq.~(\ref{wfuncany1}), we start from the normalization condition (no sum over $m$ or $k$)
\begin{eqnarray}
1 = \int {\cal D}\mu_F({\bf A}) \tr \Big{[}\psi^{[\alpha]}_{[Ad]m}({\bf A})^\dagger \psi^{[\alpha]}_{[Ad]m}({\bf A})\Big{]}= \frac{1}{2} \sum_a \int {\cal D}\mu_F({\bf A}) |\phi_{a k}({\bf A})|^2 \, ,
\end{eqnarray}
and we then construct the state projector (no sum over $k$)
\begin{eqnarray}
{\cal P}^{[\alpha]}_{[Ad]}=D_{[\alpha]}\int_{{\rm SU}{(N_c)}} {\cal D} V \phi_{a k}({\bf A}_2^V) \phi_{b k}^*({\bf A}_1^V) R_{ca}^{[Ad]}(V) R_{db}^{[Ad]}(V) \lambda^c_{s_2 r_2} ({\lambda^d}_{s_1 r_1})^*\, .\label{PRO3}
\end{eqnarray}
This expression can be simplified by recalling the transformation properties $\phi_{ak} ({\bf A}^V) = R^{[\beta]}_{kk'} (V) \phi_{a k'} ({\bf A})$ (see eq.~(\ref{wfuncany1})) and the fact that the wave-functions $\phi_{ak}$ must be such to project out the irreducible representation $[\alpha]$ from the tensor product $[\beta]\otimes [N^2_c-1]$. We get from eq.~(\ref{PRO3})
\begin{eqnarray}
&&{\cal P}^{[\alpha]}_{[Ad]}=D_{[\alpha]}\int_{{\rm SU}{(N_c)}} \!\!{\cal D} VR^{[\beta]}_{kk''} (V) \phi_{a k''} ({\bf A}_2) (R^{[\beta]}_{kk'} (V) \phi_{b k'} ({\bf A}_1))^* R_{ca}^{[Ad]}(V) R_{db}^{[Ad]}(V) \lambda^c_{s_2 r_2} ({\lambda^d}_{s_1 r_1})^*\!=\!\nonumber\\
&&=D_{[\alpha]}\int_{{\rm SU}{(N_c)}} {\cal D} V \phi_{a k''}({\bf A}_2) \phi_{b k'}^*({\bf A}_1) R_{ck;ak''}^{[\alpha]}(V) (R_{dk;bk'}^{[\alpha]}(V) )^*\lambda^c_{s_2 r_2} ({\lambda^d}_{s_1 r_1})^*=\nonumber\\
&&=\sum_{a,k} \phi_{a k}({\bf A}_2) \phi_{a k}^*({\bf A}_1)\sum_b\lambda^b_{s_2 r_2} ({\lambda^b}_{s_1 r_1})^*\, ,\label{PRO4}
\end{eqnarray}
where in the second line the matrix $R_{ck;ak'}^{[\alpha]}(V)$ is the irreducible $[\alpha]$ component in the tensor product $R^{[\beta]}_{kk'} (V)R_{ca}^{[Ad]}(V)$.

Going back to eq.~(\ref{PRO3}), we see that the contribution of this state multiplet to the ${\bf A}$ partial trace of the kernel is
\begin{eqnarray}
&&D_{[\alpha]}\int_{{\rm SU}{(N_c)}}  {\cal D} V \int {\cal D}\mu_F({\bf A}) \phi_{a k}({\bf A}^V) \phi_{b k}({\bf A}^V)^* R_{ca}^{[Ad]}(V) R_{db}^{[Ad]}(V) \lambda^c_{s_2 r_2} ({\lambda^d}_{s_1 r_1})^*e^{-E^{[\alpha]}T} = \nonumber\\
&&= \frac{D_{[\alpha]}}{N_c^2-1} 2 \sum_c \lambda^c_{s_2 r_2} {\lambda^c}_{r_1 s_1} e^{-E^{[\alpha]}T} \, .
\end{eqnarray}
One can again check that tracing also over source indices yields precisely $D_{[\alpha]}$.

\section{Decomposing the $[N_c]\otimes [\bar N_c]$ tensor product} 
\label{sec:APPC}

The $[N_c]\otimes [\bar N_c]$ tensor product representation acts on the space of complex matrices $w_{sr}$, $s,r=1,\ldots,N_c$, endowed with the scalar product
\begin{equation}
(w_2 , w_1) = \sum_{sr} {w_2}^*_{sr} {w_1}_{sr} =\tr [w_2^\dagger w_1] \, .\label{prsc}
\end{equation}
From the normalization $\tr [\lambda^a \lambda^b] = \frac{1}{2}  \delta_{ab}$, we have the identity
\begin{eqnarray}
w - \frac {I_{N_c}}{N_c} \tr [w] = 2 \sum_a \lambda^a \tr[\lambda^a w] \, ,\label{ident}
\end{eqnarray}
which implies
\begin{eqnarray}
w=\frac {I_{N_c}}{N_c} \tr [w] + 2 \sum_a \lambda^a \tr[\lambda^a w] \, , \label{ident1}
\end{eqnarray}
where $I_{N_c}$ is the $N_c\times N_c$ unit matrix. In components eq.~(\ref{ident1}) reads
\beq
w_{s_2r_2}=\sum_{s_1r_1}w_{s_1r_1}\delta_{s_2s_1}\delta_{r_2r_1}=\sum_{s_1r_1}w_{s_1r_1}\Big{(}\frac{1}{N_c}\delta_{s_2r_2}\delta_{s_1r_1}+2\sum_a\lambda^a_{s_2r_2}\lambda^a_{r_1s_1}\Big{)}\, .
\label{COMPN}
\eeq
Eq.~(\ref{COMPN}) is the algebraic identity
\beq
\delta_{s_2s_1}\delta_{r_2r_1}= \frac{1}{N_c}\delta_{s_2r_2}\delta_{r_1s_1}+2\sum_a\lambda^a_{s_2r_2}\lambda^a_{r_1s_1}\, .
\label{COMP}
\eeq
The two tensors in the r.h.s.\ of eq.~(\ref{COMP}) are precisely the two projectors onto the singlet and the adjoint representation, as it follows from the identities~\cite{Brown:1979ya}
\beqn
&&P^{[S]}= \frac{2}{N_c}\sum_a\lambda_{s_2s_1}^a\lambda_{r_1r_2}^a +\frac{1}{N_c^2}\delta_{s_2s_1}\delta_{r_1r_2}=\frac{1}{N_c}\delta_{s_2r_2}\delta_{r_1s_1}\, ,\label{NFN2}\\
&&P^{[Ad]}= -\frac{2}{N_c} \sum_a\lambda_{s_2s_1}^a\lambda_{r_1r_2}^a +\frac{N_c^2-1}{N^2_c}\delta_{s_2s_1}\delta_{r_2r_1}=2\sum_a\lambda_{s_2r_2}^a\lambda_{r_1s_1}^a\label{NFN1}\, .
\eeqn
Notice that within each irreducible representation space the two tensors~(\ref{NFN2}) and~(\ref{NFN1}) act as unit operators of appropriate dimension.

\section{The $q\bar q$-kernel in the Coulomb gauge}  
\label{sec:APPD}

For completeness we give in this Appendix the explicit expression of the Feynman kernel in the presence of external sources in the Coulomb gauge. The formulae collected in the previous sections can be straightforwardly rewritten in the Coulomb gauge because, as proved in~\cite{Rossi:1979jf}, not only the energies (that are gauge invariant quantities) but also the eigenstates of the Hamiltonian are exactly the same in the Coulomb and the temporal gauge. In fact, in going from the Coulomb to the temporal gauge, the Feynman kernel stays the same as the two gauges are related by a change of variables in the functional integral~\cite{Rossi:1979jf}.

Explicitly eq.~(\ref{KQQBAD}) takes the form 
\beqn
\hspace{-.9cm}&&K({\bf A_2},s_2,r_2;{\bf A_1},s_1,r_1;T)
=\int^{{\bf A}({\bf x},T_2)={\bf A}_2({\bf x})}_{{\bf A}({\bf x},T_1)={\bf A}_1({\bf x})}{\cal D}\mu_C({\bf A})\prod_{{\bf x}, T_1\leq t\leq T_2}\!\!dA_0({\bf x},t) \label{KQQBC}\\
\hspace{-.9cm}&&e^{-S_{YM}({\bf A}, A_0)}\Big{[}T\exp{\Big{(}i\int_{T_1}^{T_2}} A_0({\bf x}_q, \tau) d\tau \Big{)}\Big{]}_{s_2s_1}\Big{[}T\exp{\Big{(}-i\int_{T_1}^{T_2}} A_0({\bf x}_{\bar q}, \tau) d\tau \Big{)}\Big{]}_{r_1r_2}\, ,\nonumber
\eeqn
where 
\beq
{\cal D}\mu_C({\bf A})=\prod_{{\bf x}, T_1<t<T_2}\!\!\Delta_C({\bf A})\delta[\nabla{\bf A}]d{\bf A}({\bf x},t) \label{COULOMB}
\eeq
is the Coulomb gauge fixed integration measure and $\Delta_C({\bf A})$ is the corresponding   FP determinant. We note that, unlike the integration over the spatial components of the gauge field, the integration over the temporal one in eq.~(\ref{KQQBC}) is extended to include also the values at the boundary times. This is related to the fact that the temporal component of the gauge field, $A_0$, plays the role of a Lagrange multiplier which enforces the Gauss' law at any instant including the initial and final time~\cite{note2}. This is why the exponentials in eq.~(\ref{KQQBC}) do not wind up in the time direction.

Setting now as in~(\ref{DECOMP}), ${\bf A}_1={\bf A}_2={\bf A}$ and integrating over ${\bf A}$, again with the Coulomb integration measure, we conclude that the resulting quantity coincides with the expression displayed in the r.h.s.\ of eq.~(\ref{DECOMP}) from which singlet and adjoint potential can be extracted.


\begin{thebibliography} {99}
\bibitem{note1} 
In our notation the gauge coupling is absorbed in a rescaling of the definition of the gauge field.
\bibitem{note2} 
Actually one of the two boundary $A_0$-integrations can be dropped.
\end{thebibliography}
\end{document}